\documentclass[11pt]{article}
\usepackage{graphicx}
\usepackage{amsmath}
\usepackage{amssymb}
\usepackage{amsxtra}
\usepackage{adjustbox}
\usepackage{wasysym}
\title{Ontological Fluctuating Lattice Cut Off}
\author{Holger Bech Nielsen\footnote{Speaker at the  Work Shop
    ``What comes beyond the Standard Models'' in Bled.}
  \\Niels Bohr Insitutet Jagtvej 155a Copenhagen N,
  hbech@nbi.dk
}

\begin{document}
\maketitle

\begin{abstract}
  Remarkably accurate fine structure constants are calculated from  assumptions further developped from two earlier publications on
  ``Approximate SU(5)...''\cite{AppSU5} and ``Remarkable scale relation,...''
  \cite{scales}.
  In ``Remarkable scale relation,...'' we have put together a series of energy
  scales related to various
  physical phenomena such as the Planck scale , a scale, which we call fermion
  tip being a certain extrapolation related to the heaviest fermions in the
  Standard model, an approximate $SU(5)$ unification scale (without susy);
  and then we found, that as
  function of the power of an imagined lattice link length supposedly relavant
  for the scale in question, these powers are rather well linearly related to
  the logarithms of the assosiated energy scales\cite{scales}. The coincidence
  of
  these scales
  fitting a straight line is remarkable and in some cases quite intriguing.
It may in fact be taken as an evidence for Nature truly/ontologically having a
  {\em fluctuating } lattice, meaning, that the size of the links say fluctuate
  quantum mechanically and from place to place and time to time. We review
  a self-reference
  obtaining  the three fine structure constants via three theoretically
  predictable
  quanties, among which is a scale on our straight line plot, namely for
  an {\em approximate} SU(5)-like  unification (SU(5) coupling relations are
  only true in a classical approximation). Concentrating on the four energy
  scales, for which most precise numbers make sense (this is new in the present
  article), we interpolate to the
  approximate unification scale to such an accuracy, that it combined with
  the quantum corrections making the deviation from genuine SU(5) delivers
  the differences between the three inverse fine structure constants agreeing
  within errors being a few units on the second place after the comma! E.g.
  we predict the difference between the non-abelian inverse fine structure
  constants at the  $Z^0$-mass $M_Z$ to be
  $(1/\alpha_2 - 1/\alpha_3)(M_Z)|_{predict}=29.62-8.42 =21.20$, while the
  experimental difference is $29.57-8.44=21.13$ both with uncertainties of
  order $\pm 0.05$.
\end{abstract}

``When I die my first question to the Devil will be:
What is the meaning of the fine structure constant?''
    Wolfgang Pauli \cite{P78}
\section{Introduction}\label{s:intro}
%Introduction \ldots (you may use your own label names)

%with more sections to follow\ldots

%%% Her indsaetter vi slidefilen:

%\documentclass{beamer}
%\usepackage{adjustbox}
   %
\def\A{{\cal A}}
\def\B{{\cal B}}
\def\Tr{{\rm Tr}}
 \newcommand{\SO}{\mathrm{SO}}
 \newcommand{\SU}{\mathrm{SU}}
 \newcommand{\unit}{\mathrm{U}}
 \newcommand{\aat}[4]
{\tilde{A}^{#1}_{#2}\atop(({#3}),({#4}))}
%\usetheme{Berlin}

%\title{Ontological Fluctuating Lattice Cut Off  }
%\author{H.B. Nielsen\footnote{Speaker at the  Work Shop
 %   ``What comes beyond the Standard Models'' in Bled.}, Niels Bohr Institut,
  %and Colin D. Froggatt, Glasgow University
%}
%\date{``Bled''   , July , 2025}

%\section{Introduction}

We have found a rather surprising phenomenological relation between a series
of about 9 energy scales\cite{scales, Bl24, Ko24}. For each energy scale in the series
we speculate
to what power the link length $a$ of a lattice quantum field theory
- we actually in this article as the word ``ontological'' in the title
suggests that a lattice truly exist in Nature - should be raised in order
to be relevant for the scale in question, say $a^n$. Then we plot the logarithm
of the energy scales versus this power $n$ as argued for. And then the scales
come on a straight line crudely at least. Some of the scales are our own
inventions for the purpose of the present work, but at least e.g. a scale for
the string tension for strings
approximately making up the hadrons\cite{HBN,Nambu, Susskind, RS, RS2}, or
the Planck scale associated to the
Newton gravitational constant $G$ are wellknown energy scales since long.

That the different scales all give points on an approximate straight line, is
a remarkable result, even if we do not quite know, what is behind this
remarkable observation. As examples of scales we have ``inflation energy
density'' and ``infletion rate''\cite{Bellomoetal, Enquist, anintr},the scale
of see-saw neutrinoes, ``see-saw''\cite{Grimus, DavidsonI, Mohapatra, King,
  Takanishi}, the domaine walls tension \cite{odm}, and the scale of
mass of a dimuon resonace, which we want to identify as a monopole related,
say bound state of monopoles, particle \cite{dimuon, Heister}.
.

We have, however, a very
suggestive explanation
being, that there truly exist a lattice, but that this lattice is far from
a regular lattice with the same lattice constant all over. On the
contrary it is crucial for our idea, that the length of the links (= the
lattice constant roughly ) varies dramatically from place to place and
we would also assume from time to time, and really we would even like the
lattice-like structure
to be in superposition of states with various link sizes, if that makes sense.
The crucial point is,
that looking at some place at some moment, one
can by accident find the lattice-like structure there very tight or
very rough as it can happen. Especially the link size $a$ will statistically
fluctuate wildly. On the figure you can see, how such a
``lattice'' which has different tightness in different places
may look in an accidental moment.

%\begin{document}
%\section{Intro.}
\begin{frame}
  
  {\bf \huge Ontological Fluctuating Lattice Cut Off}

%  \maketitle
\end{frame}
\begin{frame}
  
  {\bf An irregular lattice with big density differences}
  
  \includegraphics[scale=0.5]{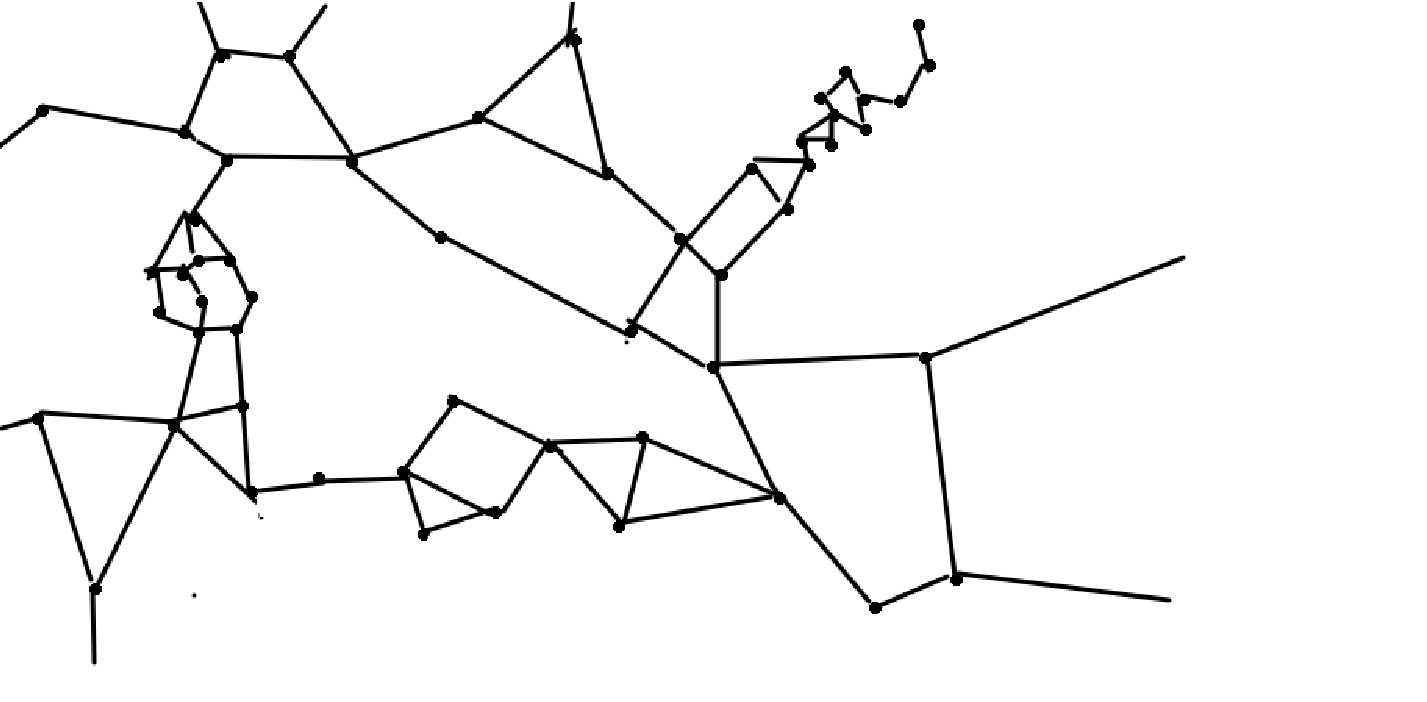}
\end{frame}

The present work may be considered a work on our
project Random Dynamics\cite{Volovik, Picek, Polonica} and for instance
some works on why just the Standard Model group \cite{sr, srKorfu14, srcim4}
and thoughts upon the possibilities\cite{ Rugh} for using numerology.
Old reviews of our Random Dynamics  are \cite{Teitelbaum, book,PS } and the
original part of an article
see\cite{RDori} comming up with the idea. The present work began by the
finestructure constant
calculations discussed in section \ref{Finestructure} below, which really
has the philosophy, that the approximate SU(5) relations between the (running)
couplings at some scale - which call in our model $\mu_u$ - is an accident
occuring 
because the classical lattice action happens to be SU(5) invariant, because
it is a representation of the Standard Model group, that happens to be a unitary
$5\times 5$ matrix. So we predit quantum corrections to SU(5), and thus true
SU(5) GUT 
 is only {\bf approximate}. But even though we thus do not have genuine
SU(5), the GUT SU(5) earlier works are used by us \cite{GG, WGG3, GG2},
but we actually in our model as a further speciality rather have
 a cross product
 of three (approximate) SU(5)'s \cite{LD, flipped, LRSU5}. Well, we should
 rather say we assume that there truly is a cross product of three copies of
 the Standard Model group with their Yang Mills fields, while the remaining
 Yang Mills components of the SU(5)'s do not exist.
 
 Especially the finestructure project
in RandomDynamics may be found in\cite{RDrel1,RDrel2, RDrel3,RDrel4, Don93,
  Don137, Bled8, crit,LRN}. The idea that the gauge groups we see should be
diagonal subgroups of say a cross product of three of the same group, was
favoured by, that we invented a mechanism ``confusion'' \cite{confusionetal,
  Mizrachi} that tended to make diagonal subgroups only survive, when possible.
So whatever the guage group behind, we should only see the diagonal subgroup,
and thus we claimed it likely, that we have several isomorphous cross preduct
factors in the true group behind. We called this anti-GUT. In any case we
assume in the present article, that the Standard Model group is repeated three
times in this way in the present article. This assumption is most relevant
for our fine
structure constants calculation.

\subsection{Our model}
\subsubsection{Finestructure Constants}

Let us resume shortly, what our model, especially for the finestructure constant
part, is: There exists a fundamental gauge theory lattice, which is fluctuating
in the sense of being somewhere and sometime very tight and somewhere and
sometimes very rough. It is for a gauge group Standard model group cross
producted with itself to third power. So each family of Fermions can have its
own out of the three cross product factors to couple to. One could equivalently
take it, that there were three lattices on top of each other, three layers so
to speak with only the Standard model group each. Such an Anti-gut is supposed
to break down to only the Standard model group (diagonal breaking)
presumably at the scale of the local lattice size. It is this breaking
we told in the old papers were due to ``confusion''
\cite{confusionetal, Mizrachi}. 
%11\%

A crucial point for the approximate SU(5) is, that we think of
the link objects as $5 \times 5$ unitary matrices, that though are
restricted to only take such values that they occur in the five dimensional
representation of the Standard model group. But thinking on the
link-variables this way makes it natural, that the plaquette actions
should be the trace of such a $5 \times 5$ matrix. But this is exactly
the same as the most natural SU(5) theory action. Since the finestructure
constants at the scale in question, the scale of the link length say, is
given by the form of the action we have in first approximation
SU(5) related couplings even though we only have the gauge group
being the Standard model one. However, now there are quantum
corrections, which are essentially the effect of quantum fluctuatons of say
 the plaquette variables, and they can in our model only be in the
directions of the really existing Standard model group, while there can be
no fluctuations in the non-existing 12 only SU(5) dimensions of
fluctuations. But now when there in our model is  one Standard model group for
each family of fermions (i.e. 3) we get the quantum correction multiplied by 3.

The most remarkable progress of the present article is, that we use the
straight line with the energy scales versus the power to which the
link length $a$ is raised in connection with the energy scale in question
to predict the replacement for the unification scale for the
approximate minimal(i.e. without susy) SU(5) to predict the differences
between the three inverse fine structure constants with the phantastic
accuracy of having the deviation from the experimental differences only
of the order 0.05 for differences that can be of order 20. I.e. less than
a percent deviation!
%13\%

We also do predict the genuine fine structure constants themselves, but
by an uncertainty rather only 3 units in the inverse fine structure constants.

It should be mentioned that Senjanovic and Zantedeschi \cite{Senjanovic}
have sought to recsue SU(5) GUT by means of higher dimensional
operators, and get to a very similar replacement for unification
as we do $10{14} GeV$. In principle our approach is very similar, since
having a lattice at least in principle could induce higher dimensional
operators.

\subsubsection{Diagonal Subgroup Fine structure constant formula}
To understand the factors 3 occuring in our formulas one should have in mind
that when we have breaking of the symmetry from a group to some
cross product power, say $G\times G \times G$ (this is a thrid power of $G$),
then if the fine strucutre constants for the here three groups isomorpkic to
$G$ are say $\alpha_{Peter}, \alpha_{Paul},\hbox{ and } \alpha_{Maria}$, then the
finestructure constant for the diagonal subgroup $G_{diag}$  of the cross product
\begin{eqnarray}
  G_{diag}&\subset&G\times G\times G\\
  \hbox{i.e. } G_{diag}&=& \{ (g,g,g)|g\in G\}
\end{eqnarray}
we have the fine structure constant relation
\begin{eqnarray}
  \frac{1}{\alpha_{diag}} &\approx & \frac{1}{\alpha_{Peter}}+
  \frac{1}{\alpha_{Paul}}
      + \frac{1}{\alpha_{Maria}}\label{diag}\\
        \hbox{or } &\approx& \frac{3}{\alpha_{Peter}},\hbox{ if }
          \alpha_{Peter}=\alpha_{Paul}=\alpha_{Maria}.
  \end{eqnarray}
\subsubsection{The Series of Scales}

The assumptions needed to make the lattice constant associated with the
energy scales, is that the coupling parameters in the lattice action are of
order unity, and in addition the assumption about Log Normal distribution
of the link variablein the fluctuating lattice, as to be described in
next section
\ref{LogNormal}.

The idea of a strongly fluctuating lattice may be supported by the
idea, that the metric of general relativity fluctuates corresponding to
diffeomorphism transformations. Such fluctuations of the metric would be
suggested, if we believe as Ninomiya F{\o}rster and myself and Shenker
suggested that gauge symmetries come about due to strong quantum fluctuations
in the gauge \cite{FNN, trans, RS}; the idea is that it is the lattice that
is part of the gauge variable, which are taken to exist. Thus we would see it
as the lattice fluctuating relative to us, although it might really rather
be the metric giving space-time, that fluctuate relative to the lattice.. 

\subsection{Plan of article}
The introduction was section \ref{s:intro}.

In the following section \ref{LogNormal} we introduce the statistical
distribution to desribe the probability distribution, if one takes
out a random link and looks fro it geometrical size $a$.

In section \ref{ThreeScales} we use the three lowest energy scales as
an example to give an idea of how we estimate the relevant power $n$ to
which the link gets raised  concerning the scale in question.

The presentation of the main plot with the points, that should by our
model have about 9 points for nine scales lie on a straight line in section
\ref{MainPlot}.

One of our energy scales, which we invented ourselves as so many of them,
is the energy scale for masses of monopoles, as we want to predict
(although these monopoles are not very strongly tied to our model,
we would prefer them to be there, but really our model could still
survive even if there did not exist monopoles), and we assign a short
section \ref{Monopole} to such monopole related particles, and we seek to
identify a particle that is not itself a monopole but rather some
by gluons confined set of monopole like particles, presumably
with no genuine monopole field around  in the outside. But it could be bound
from particles with genuine monopolic magnetic charge. We seek to identify
this monopole related particle with what is presumably a statistical
fluctuation peak in $muon^+ + muon^-$ at a mass 27 GeV at LHC, but
also seen at LEP.

A last section \ref{Finestructure} before the conclusion is assigned
to our prediction of the finestructure constants for the three simple
Lie algebras of the Standard
Model.

In section \ref{Conclusion} we conclude and make a tiny outlook.

\section{Log Normal Distribution}
\label{LogNormal}
The Log Normal distribution\cite{LN} (sometimes called after Galton, McAlister,
Gibrat, or Cobb-Douglas)
means a statisical distribution of a
quantity, such that the logarithm of this quantity becomes distributed
as a normal distribution around some mean, but  with the
Gaussian distribution being for the  logarithm. Such a Log Normal
will, in analogy to the central limit theorem for summing a lot of
(only weakly dependent) random variable, result, if one mutiplies a lot
of random variables. E. g. imagine a human being playing with his fortune
on the bourse or making speculations some way or another, then typically
the gain or loss will be proportional to the amount of money, with which he
has been able
to speclulate, and thus to his fortune. After many such speculations
we should be able to trust that the probability distribution for his fortune
- at the late time - should have a Log Normal distribution.
Indeed the distributions of fortunes of different poeple on the Earth
is rather well a Log Normal one.
If the lattice,
Nature has given us (remember in this article we believe there truly exist
a lattice, ontological lattice),has it so that, if we go around in it the link
size will locally  vary by some not too big factor up or down, with about the
same probability, independently of whether the lattice locally is tight or
rough, then the lattice link distribution will end up roughly a Log Normal
distribution also.

We like to think of the Log Normal distribution as a distribution, which under
very mild assumptions comes by itself. It is really like it should be in
Random Dynamics, that even say if the truth were not a lattice but some other
type of cut off- a cut off momentum-energy scale $\Lambda$ say - then we
should still guess a Log Normal distribtuion for this other type of cut off
parameter, $\Lambda$ say.

\begin{frame}

\subsection{{\bf Main Philosophy}}
%\vspace{-2mm}
The main point of our work is to assume, that we have a lattice
- this shall then be fluctuating in tightness, being somewhere tight,
somewhere rough with big links and net holes - and then the various
physical energy scales are calculated each of them from some power of the
length $a$ of a link. While for a rather narrow distrbution of a variable
$a$ say it is so that whatever the power of the variable $a$ you need for your
purpose you get about the same value for the effective typical $a$ size,
\begin{eqnarray}
  \sqrt[n]{<a^n>} &\approx& \hbox{ $n$-independent} \hbox{ (for narrow
    distributions.}\nonumber\\
  \hbox{However, Galton}&& \hbox{distribution: }\nonumber \\
  P(\ln a)d\ln(a) &=&
  \frac{1}{\sqrt{2\pi \sigma}}\exp(-\frac{(\ln a - \ln a_0)^2}{2\sigma}) d\ln a
  \label{Galton}\nonumber\\
  \hbox{gives rather }\sqrt[n]{<a^n>} &=& a_0\exp(\frac{n}{2} *\sigma).
  \label{ese}
\end{eqnarray}

  \end{frame}
\begin{frame}

  {\bf Exceptional case $n=0$:}

  The expression $\sqrt[0]{<a^0>}$ is not good but we reasonably
  replace it
  \begin{eqnarray}
    \sqrt[0]{<a^0>} \rightarrow \exp(<\ln(a)>) &=& a_0 \\
    \hbox{for our Log Normal.} P(\ln a)d\ln(a) &=&
    \frac{1}{\sqrt{2\pi \sigma}}\exp(-\frac{(\ln a - \ln a_0)^2}{2\sigma})
    d\ln a\\
\hbox{so again } \sqrt[n]{<a^n>} &=& a_0\exp(\frac{n}{2} *\sigma)
  \end{eqnarray}
  \end{frame}
\begin{frame}

  \subsection{Fluctuating Lattice stressed}
  
{\bf Comparing Our fluctuating lattice with usual Wilson one}
  
  \includegraphics[scale=0.5]{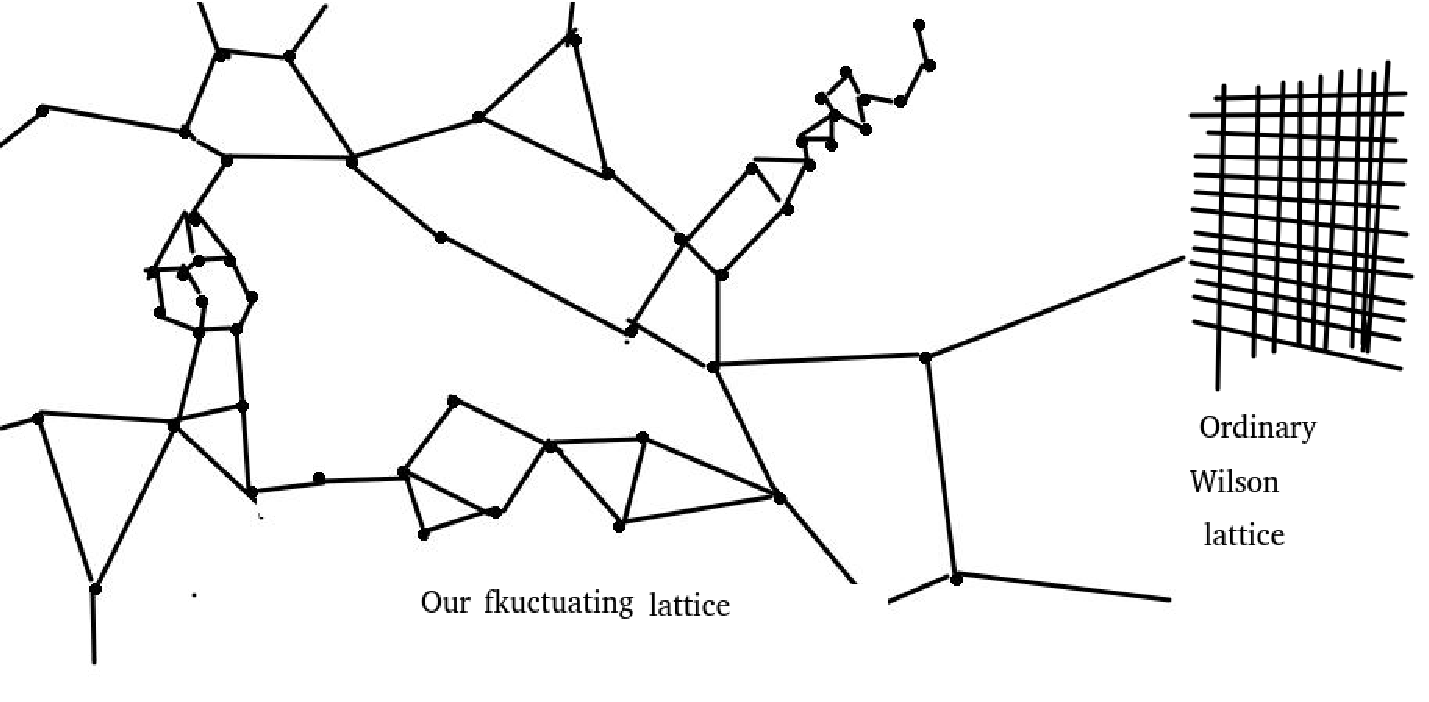}
  \end{frame}

\begin{frame}

  {\bf Ontological lattice mean really in Nature existing lattice}
\begin{center}
  {\bf Usual non-fluctuating lattice}$\rightarrow$ {\bf Fundamental scale}.
  \vspace{1 mm}
  
  {\bf Our fluctuating lattice} $\rightarrow$ {\bf Several different fundamental scales.}
\end{center}
  
\end{frame}
\section{Example of Three Scales}
\label{ThreeScales}
\begin{frame}

  {\bf Introductory Examples of Powers of the Link Size to Average}
  
To get an idea of how we may derive the relevant average of a power $<a^n>$
let us
for example think of a particle, a string, or a domaine wall being
described by action of the Nambu-Goto types
%\begin{subequations}
\begin{eqnarray}
  \hbox{Particle action} S_{particle} &=&
  C_{particle}\int \sqrt{\frac{dX^{\mu}}{d\tau}g_{\mu\nu}\frac{dX^{\nu}}{d\tau}}
  d\tau\\
  &=& C_{particle}\int \sqrt{\dot{X}^2}d\tau\label{particle} \\
  \hbox{String action }S_{string} &=& C_{string}\int d^2\Sigma \sqrt{(\dot{X}\cdot
    X')^2 - (\dot{X})^2(X')^2}\\
  &=& -\frac{1}{2\pi \alpha'}\int d^2\Sigma \sqrt{(\dot{X}\cdot X')^2-
    (\dot{X})^2(X')^2}\label{string}
\end{eqnarray}
\end{frame}

\begin{frame}

  {\bf Domain wall action}
  
\begin{eqnarray}
  \hbox{Domain wall action } S_{wall}&=& C_{wall}\int d^3\Sigma \\
  &&{
  %\left
    \det{  \begin{bmatrix} (\dot{X})^2 & \dot{X}\cdot X'&
      \dot{X}\cdot X^{(2)}\\
      X'\cdot \dot{X}& (X')^2 & X'\cdot X^{(2)}\\
      X^{(2)}\cdot \dot{X} & X^{(2)}\cdot X' & (X^{(2)})^2
    \end{bmatrix}}
    %\right
    }\label{wall}
  \end{eqnarray}
%\end{subequations}
Here of course these three extended structures are described by
repectively 1, 2,
and 3 of the parameters say $\tau, \sigma, \beta$, the derivatives with
respect to which are denoted by respectively $\dot{ }, ', and {}^{(2)}$. So
e.g. $d^3\Sigma = d\tau d\sigma d\beta$ and
\begin{eqnarray}
  X^{(2)}&=& \frac{\partial X^{\mu}}{\partial \beta}\\
  X'&=&\frac{\partial X^{\mu}}{\partial \sigma}\\
  \dot{X} &=& \frac{\partial X^{\mu}}{\partial \tau}\\
  \end{eqnarray}
Finally of course $\cdot$ is the Minkowski space inner product.

Now imagine, that in the world with the ontological lattice, which
we even like to take fluctuating, these tracks of objects, the
particle track, the string track or the wall-track, should be identified
with selections of in the praticle case a series of links, in the string case
a suface of plaquettes, and in the wall-case a three dimensional structure
of cubes, say. One must imagine that there is some dynamical marking of the
lattice objects - plaquettes in the string case e.g.- being in  an extended
object.  Now the idea is that we assume the action
%of the lattice to be
%involving interactions etc.
for the lattice to have 
%with the
parameters
%being
of order unity.
In that case the order of magnitude of the effective tensions meaning the
coefficients $C_{particle}, C_{string}, C_{wall}$ can be estimated in terms of
the statitical distribution of the link length - for which we can then
as the ansatz in the model take the Galton distribution (\ref{Galton}) -
%26\%
by using respectively the averages of the powers 1, 2, 3, for our three
types of extended objects. I.e. indeed we say that by order of magnitude,
the mass of the particle, the square root of the string energy density
 or the string tension, and the cubic root of the domain wall tension
are given as the inverses of averages of $a$ like:
\begin{eqnarray}
  \hbox{Particle mass } m &\sim & <a>^{\; -1}\label{Pm}\\
  \hbox{Square root of string tension } \sqrt{\frac{1}{2\pi \alpha'}} &\sim &
  \sqrt{<a^2>}^{\; -1}\label{Stsr}\\
  \hbox{String tension itself } \frac{1}{2\pi \alpha'} &\sim& <a^2>^{-1}\\ 
  \hbox{Cubic root of wall tension } S^{1/3} &\sim& \sqrt[3]{<a^3>}^{\; -1}
  \label{Wtcr}\\
  \hbox{wall tension itself } S &\sim& <a^3>^{-1}.
  \end{eqnarray}
Here the $\sim$ approximate equalities are supposed to hold order of
magnitudewise under the assumption that no very small or very big numbers are
present in the coupling parameters of the lattice, so that it is the somehow
averaged lattice that gives the order of magnitude for these energy densities
or tensions.

\subsection{Illustration of the idea}
Although our speculations for the three energy scales - meaning numbers with
dimension of energy - which we in my speculation attach to these three
objects, the particle, the string, and the wall, are indeed very speculative
only,
and that we shall give a bit better set of such scales in next subsection, let
us nevertheless as a pedagogical example consider these three first:

\end{frame}
\section{Main Plot}
\label{MainPlot}
\begin{frame}
  
  {\bf Our Phantastic Plot: Agreeing Order-of-magnitudewise for 9 Energy
    scales} 
\begin{figure}
  \includegraphics[scale=0.3]{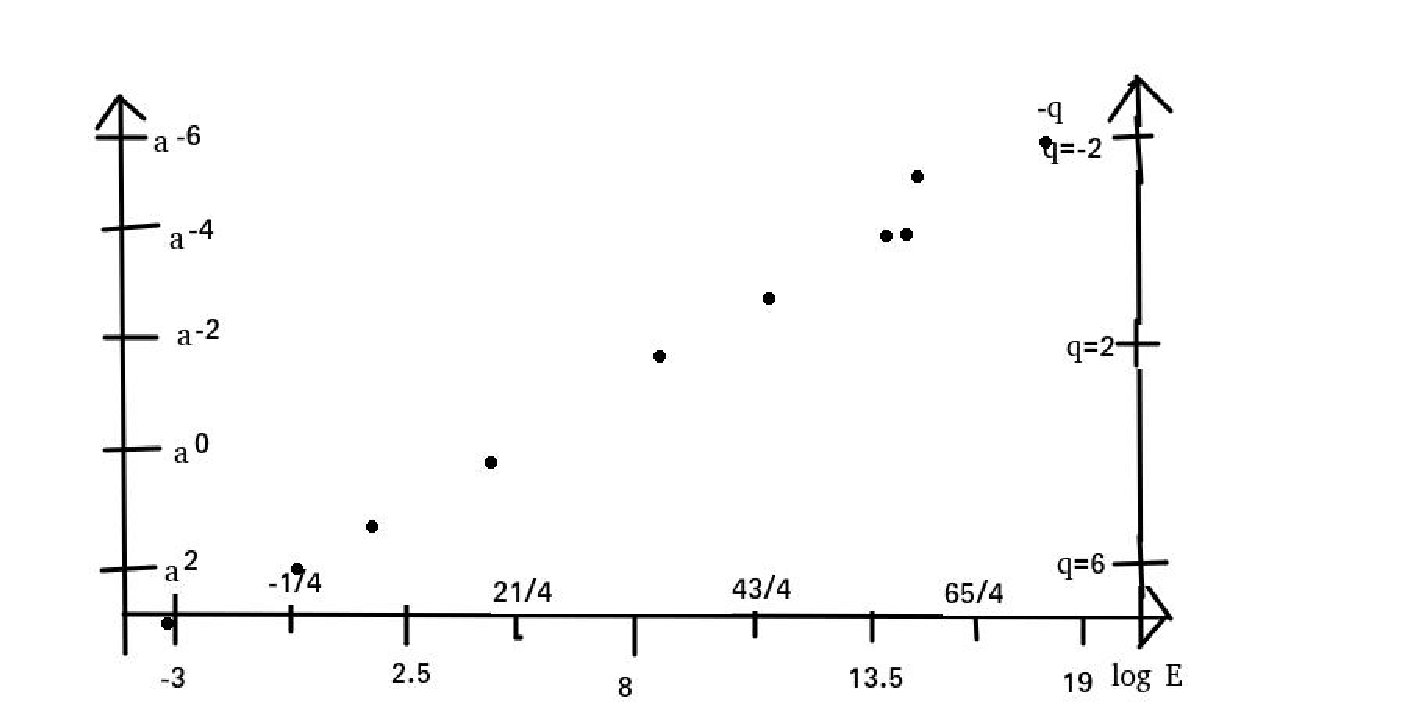}
  \caption{\label{wd}
    %Similar figure as
    %figure \ref{all} but only with reduced Planck scale
    %for the Planck scale, and written with just dots to let us
    %enjoy how well it fits the straight line. The cosmological entries
    %looks worst and susy grand unification was left out.
    Plot of the (inverse) power $n$, into which comes the lattice link
    length $a$,
    when forming the physical energy scale of energy $E$, versus its  logarithm
  of this energy with basis 10 $log \; E$ using GeV as energy unit}
\end{figure}
\end{frame}
\begin{frame}

  {\bf Main Plot a bit bigger and with names}
  \vspace{-6 mm}
  
  \hspace{-6 mm}\includegraphics[scale=0.55]{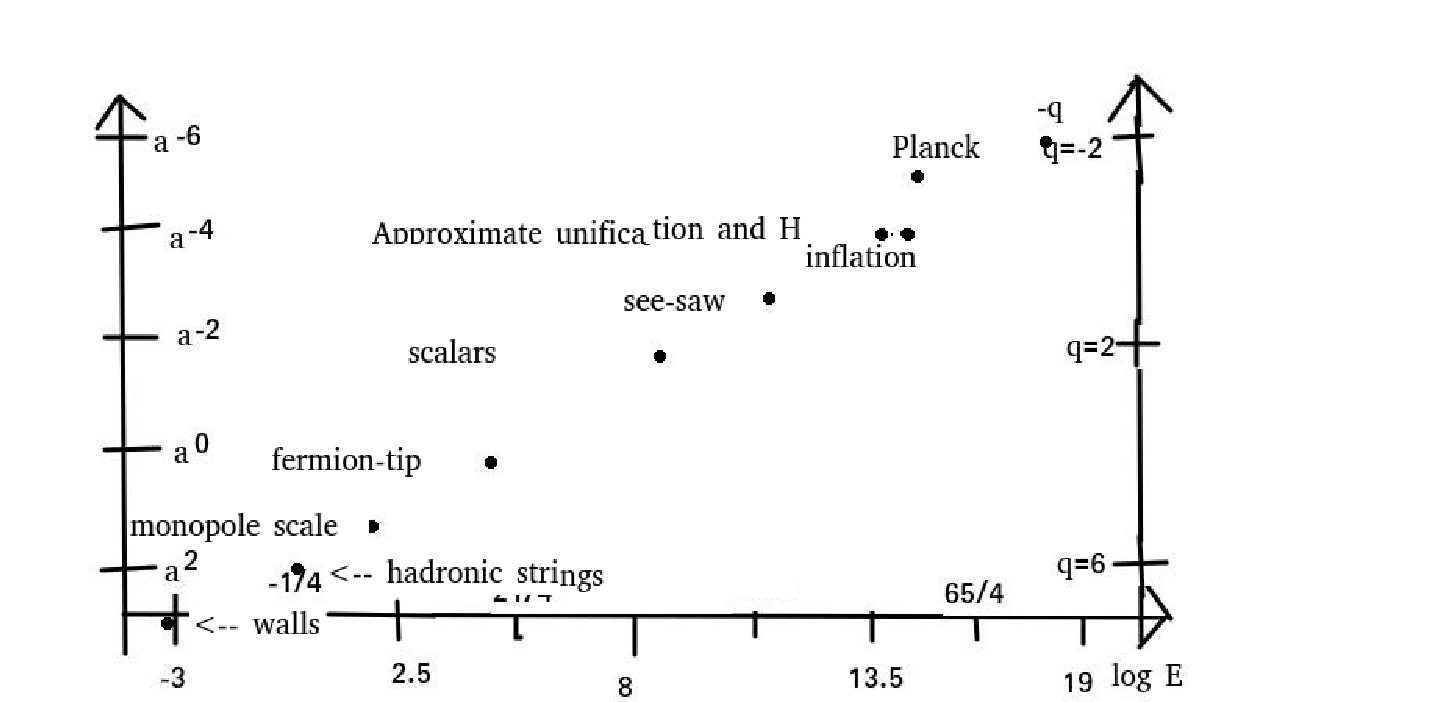}
\end{frame}

\begin{frame}

  {\bf Comments on the Main Plot}
  \begin{itemize}
  \item Two cosmological points ``$H_{inflation}$'' and one related to the
    energy density in the inflation time (which I did not give a name, but
    call it $\sqrt[4]{V_{inflation}}$), do not fit quite perfect on the straight
    line, but not so badly either.
  \item For PlanckscaleIused the reduced planck scale with an 8$\pi$
    divided out of the gravitationalconstant $G$. But this theoretically best
    suggested.
  \item Some energy scales are wellknown: ``Planck'', the cosmological ones,
    an only approximate SU(5) ``unification'' scale, The ``see-saw'' mass scale,
    the energy scale of hadron  physics taken as an approximate ``string''
    theory
    i.e. $\alpha'$ defining a scale.
  \end{itemize}
\end{frame}

\begin{frame}

 % {\bf Comments continued, Now the speculated scales}
  \begin{itemize}
  \item But the rest is my own inventions/speculations:

    {
      %\color{red}
      {\bf ``scalars''}} meaning
    I speculate that a lot of scalars have their mass and possible vacuum
    expectation values of this energy order of magnitude;

    {
      %\color{red}
      {\bf ``fermion tip''}}
    which is the tip or top of an extrapolation of the density of the numbers
    of standard model fermion masses on a logarith of the mass abscissa;

    {
      %\color{red}
      {\bf ``monopoles''}} a
    certain dimuon resonance barely observed of mass 27 GeV is speculatively
    taken as being somehow to monpoles, which though are presumably cofined
    because of their QCD features;

    {
      %\color{red}
      {\bf ``walls''}} are the domain walls around
    dark matter in my own and Colin Froggatts darkmatter model, their
    energy scale is the third root of the wall tension.
    \end{itemize}
  
\end{frame}

\subsection{Remarks on Higgs Etc.}

In the second paragraph of the introduction, section \ref{s:intro}, we
mentioned several  of the energy scales, which we consider, but did not
mention that very speculatively we predict an energy  scale given
by $\sqrt{<a^{-2}>}$, at which there should be ``a lot'' of scalar
particles having their mass - analogous to that the see-saw scale is a
scale at which ``a lot''(at least there must be one or two) fermions have
their mass -. Now it would have been very nice for our model, if
the Higgs mass had been at this scale, but that is not at all the case.
It is wellknown that the small Higgs mass is a mystery, and that there is
even the hierarchy problem that the bare Higgs mass must be surprisingly much
finetuned. The author actually has a very different story about 
why the Higgs mass
should be so small, based on the idea that there is a complex action
and a selection principle of the initial conditions (the details of the
cosmological start) so as to arrange many things, the future and probably
even the parameters like the Higgs mass, so as to minimize the Higgs field
square\cite{Coincidensies, Relation, Higgsmass}. But this is quite different
story than the present article
\cite{CAT2, CAT}, but it means that the present article had no luck with the
Higgs mass.

However, this energy scale with the lot of scalar masses, would expectedly
also lead to some non-zero expectaton values in vacuum for these scalar fields
and thus break symmetries spontaneously. This we claimed in the previous article
as a good candidate for the breaking of the symmetries leading to the rather big
mass ratios of the quarks and the charged leptons. The order of magnitudes for
this ``little hierarchy''\cite{sf,qarketcmasses,Ferrucchio} mass ratio
problem is actually fitting with the ratio
of the see-saw scale to our invented and predicted many scalars
scale\cite{scales}. 
%32\%

\section{``Monopole Energy Scale}
\label{Monopole}
%\begin{frame}

  When we have gauge group $SMG=S(U(2)\times U(3))$ suggested in the
  O'Raifeartaigh-way \cite{OR} \footnote{O'Raifeartaigh propose
    to assign a {\bf group} and not only a Lie algebra to a phenomenologially
    found model, such as the Standard Model, by using the information
    about the representations found to exist physically and choose
    that Lie {\bf group} which allows as few as possible representations,
    but nevertheless the physically realized ones. On a lattice then
    the link variables and thus also the plaquette variables should
  run on/take values in  this chosen {\bf group}.} on a lattice one gets monopoles
  unless it is the covering {\bf group} (i.e. a simply connected group). We thus
  do expect monopoles, if we
  take the lattice seriously, in fact monopoles coresponding the
  the three copies of the Standard Model group $SMG$ assumed in our model
  in the cross product. Because of subgroup of the center of the
  covering group divided out involving all the three groups U(1), SU(2),
  and SU(3), the monopoles will have magnetic fields from all the three
  groups. Especially they would have gluon fields around them, and
  it is easy to imagine them getting clumped together by the confining
  vacuum(of QCD).

\begin{figure}
    \caption{\label{fig2}
  {\bf Can this Peak in $\mu\bar{\mu}$ be Monopole Related?}}

  \includegraphics[scale=0.85]{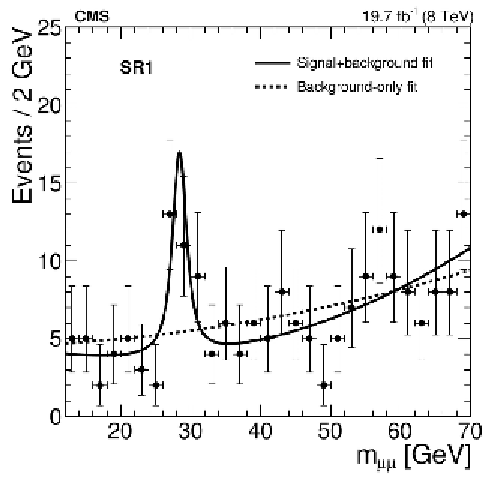}
%\end{frame}
\end{figure}

  If we really shall associate the dimmuon resonance perhaps observed
  \cite{dimuon, Heister}(see figure \ref{fig2}) with monopoles, of course it must be a combination of
  monopoles with
  zero monopolic charge. A true monopole would be stable of course.
  But a bound state living long enough to be seen as a resonance is not
  excluded.

  Indeed a dimuon decaying has been seen in events selected with some
  b-activity in LHC\cite{dimuon}; but most remarkably Heister could dig a
  similar
  resonace out of the data of already stopped LEP\cite{Heister}.
  %\newpage

\section{The four best points on our line}
\label{FourBest}
Many on the points on our straight line, being our great success, are only
definable order of magnnitudewise, because they involve thoertical models
not yet
well established. Such lack  of even more than order of magnitude numbers
due to there being no established theory is the trouble for the energy
scales, such as
the ``see-saw'', the inflation connected scales $H_{inflation}$,...,and
  the ``monopole scale''. For our ``domaine wall'' scale\cite{FNdm} the uncertainties
  in the parameters are so big, that even trusting our theory of  dark
  matter, the energy scale would be ill defined, and known to of order of
  magnitude at
  best. But for {\bf four of our scales} it is at least possible to write down
  some more accurate numbers, and if we postulate, that we shall use the
  appropriate root of the quantity occuring as a coefficient in the Lagrangian,
  we can argue, that we shall chose the reduced planck scale and
  we make at least formally a welldefined number for the Planck scale.
  Similarly the quantity $\frac{1}{4\pi \alpha'}$ occuring in the 
  string action could give a welldefined number for the  ``string-'' scale.
  When looking for an approximate unification
  or replacement for the unication scale,the value of the energy scale
  $\mu_u$ is of course renormalisation scheme dependent, but that we can hope
  is not causing too much uncertainty in practis. The fermion tip scale is
  in priciple an extrapolation
  from masses which are well measured.
  %35\%
  
  \subsection{String scale}
  In the string theory for hadrons\cite{HBN,Nambu,Susskind}, which never
  became perfectly working,
  the coefficient in the action for the string is given by
  $\frac{1}{2\pi \alpha'}$, where $\alpha(m^2)$ is the Regge trajectory
  as function of mass square.
  %where the slope parameter $a$ essentially
  %&$a=1/\alpha'$ is the slope of the function of the mass of the quark
  %with an $mass^2$ as the important term for us. 
  S. S. Afonin and I. V. Pusenkov\cite{RS} find in their fit I, that
  $1/\alpha'=1.10 GeV^2$, while Sonnenshein et al. \cite{RS2} find
  $\alpha' = 0.95 GeV^{-2}$ and looking at various Chew Frautschi plots\cite{CF}
  you see e.g. the $\rho$ and $a$ pair slope of $m^2$ versus spin $J$ means
  $1/\alpha'= 1.16 GeV^2$, for $\omega$ and $f$ almost the same, and for
  $\phi$ $f'$ a slope
  $1/\alpha' = 1.2GeV^2$. As a midle between the large $1/\alpha'$ from
  Chew Frautischi plots
  by eye and the small $1/\alpha' = 1.05 GeV^2$ from Sonnenshein et al
  \cite {RS2} we may choose the Afonin et all $1/\alpha' = 1.10 GeV^2$:  
  \begin{eqnarray}
    \hbox{Slope } a&=& 1.10 GeV^2 \hbox{(Afonin et al.)}\\
    \hbox{Thus } \alpha' &=& 1/a =0.909 GeV^{-2}\hbox{($a$ is here not the
      link length)}\nonumber\\
    \hbox{and }\frac{1}{2\pi \alpha'} &=& \frac{a}{2\pi}=0.175 GeV^2 \\
    \hbox{so that} ``string\;  scale''&=& \sqrt{0.175 GeV^2} = 0.418 GeV\\
    \hbox{and } log(`` string\;  scale'') &=&-.3784 (\hbox{in GeV}) 
  \end{eqnarray}

  %38\%
  \subsection{Fermion tip scale}

  A few estimates, 5  fermion masses are combined with the top-mass
  to give a value for the extrapolation to the ``fermion tip'' below
  here (the combinations not involving the top mass are not
  used in our averaging to get a most accurate ``tip'' value):
  \begin{eqnarray}
    &\hbox{From b to t}&\nonumber\\
    \frac{(2.2374 -0.6212)\sqrt{3.5}}{\sqrt{9.5}-\sqrt{3.5}}+2.2374&=&4.7334\\
    10^{ 4.7334} &=& 5.4129*10^4 GeV\\
    &\hbox{From $\tau$ to t}&\nonumber\\
     \frac{(2.2374 -0.2496)\sqrt{3.5}}{\sqrt{13.5}-\sqrt{3.5}}+2.2374&=&4.2995\\
     10^{ 4.2995} &=& 1.9930*10^4 GeV\\
     &\hbox{From $\tau$ to b}&\nonumber\\
     \frac{(0.6212 -0.2496)\sqrt{9.5}}{\sqrt{13.5}-\sqrt{9.5}}+0.6212&=&
     2.5558\\
     10^{ 2.5558} &=& 3.5959*10^2 GeV\\
     &\hbox{From c to t}&\nonumber\\
     \frac{(2.2374 -0.1038)\sqrt{3.5}}{\sqrt{17.5}-\sqrt{3.5}}+2.2374&=&3.9635\\
     10^{ 3.9635} &=& 9.1942*10^3 GeV\\
      &\hbox{From $\mu$ to t}&\nonumber\\
     \frac{(2.2374 -(-0.9761))\sqrt{3.5}}{\sqrt{21.5}-\sqrt{3.5}}+2.2374&=&
     4.4109\\
     10^{4.4109} &=& 2.5758*10^4 GeV\\
      &\hbox{From s to t}&\nonumber\\
     \frac{(2.2374 -(-1.0292))\sqrt{3.5}}{\sqrt{25.5}-\sqrt{3.5}}+2.2374&=&
     4.1598\\
    10^{4.1598} &=& 1.4448*10^4 GeV\\
    \end{eqnarray}

  \subsection{Some averaging}

  The average of the fermion tip scale calculated ``from the uppermost 5
  fermions (except for t) to t'' gives
  \begin{eqnarray}
    log( \hbox{`` fermion tip'' }_{first \; 5})&=& \frac{4.7334+4.2995+
      3.9635+4.4109+4.1598}{5}\nonumber\\
    &=& 4.31342\\
    10^{4.31342}&=&2.0578*10^4 GeV 
  \end{eqnarray}
  
  \subsection{Resume the Four Scales with Precise Numbers}

  The numbers obtained for the four scales, for which we can meaningfully
  write more than order of magnitude numbers are
  \begin{eqnarray}
    \hbox{``string scale''} = 0.418\pm 3\% GeV&\rightarrow& log = -0.3788\pm
    0.013\nonumber\\
    \hbox{``fermion tip''} =2.0578*10^4GeV\pm 60\% &\rightarrow&
    log = 4.3134\pm 0.2\\
    \hbox{``unification scale''(R)}=\mu_u = (5.116\pm 0.1) *10^{13}GeV
    &\rightarrow& log = 13.7090 \pm 0.01\\
    \hbox{``unification scale''(D)}=\mu_u = (4.383\pm 0.1) *10^{13}GeV
    &\rightarrow& log = 13.6419 \pm 0.01\\
    \hbox{``Reduced Planck scale''}= 2.434*10^{18}GeV\pm 2.2*10^{-3}\%
    &\rightarrow& log = 18.3862 \pm 0.000001\nonumber
  \end{eqnarray}

  (The two by 14\% different scales given for the unification
  $\mu_u$ are obtained from our model, in which the deviation from true
  minimal $SU(5)$
  GUT can be considered a quantum correction, by requiring respectively
  \begin{itemize}
  \item{{\bf R:}} that the ratio of the differences among the three
    $1/\alpha_i(\mu_u)$ be the right one;
  \item{{\bf D:}} that the outermost of the three $1/\alpha_i(\mu_u)$'s
    have the right difference $q$)
\end{itemize}

%41\%

  The power of the link variable $n$ in the relevant $\sqrt[n]{<a^n>}$
  for the string scale and the Reduced Planck scale are respectively
  $n= 2$ and $n=-6$. Thus the difference is 8, and the increase in
  the energy scale per decrease in the value of $n$ is a factor
  \begin{eqnarray}
    \hbox{``factor''}=\exp(\sigma/2)&=& \sqrt[8]{\frac{2.343*10^{18}GeV}{0.418
        GeV}} = 220.584\\
    \hbox{or } log(\hbox{``factor''})=log(e)*\sigma/2 &=&
    \frac{18.3862-(0.3788)}{8}
      \\
      &=&18.7650/8=2.345625 \\
      \sigma/2 &=& \ln(220.584) =5.3963\\
      \hbox{Width in ln: } \sqrt{\sigma} &=& 3.285
  \end{eqnarray}
  (Our notation in the present article is so, that we used $\sigma$ for what
  is ususally called $\sigma^2$, so that one standard deviation in the
  logarithm is actually given by our $\sqrt{\sigma} =3.285$ meaning, that
  since the average is at the logarith of the ``fermion tip'' scale
  at $2.06 *10^4 GeV$, the one standard deviation scales are
  $2.06*10^4GeV*\exp(\pm 3.285)$ and becomes increase and decrease
  by $26.71$. One standard deviation under the average
  reaches down to $771.3$ GeV)
  
  The slope ``factor'' obtained using `fermion tip'' together with the
  Planck scale instead of the string-scale is (accidentally ?)
  very close to the already obtained slope. The value for the unfication
  scale predicted by our
  straight line is
  \begin{eqnarray}
    \mu_u &=& 10^{18.3862-2*2.345625}=10^{13.6949}\\
    &=&4.9534*10^{13}GeV.\label{muupred}\\
    \hbox{this means } \ln(\frac{\mu_u}{M_Z})& = & \ln(\frac{4.9534*10^{13}GeV}
         {91.1876 GeV})\\
         &=& \ln(0.05432) +13*\ln(10) =27.0208.\label{lnmuuomz}
    \end{eqnarray}
  It falls between the two by different requirements methods ({\bf R} and
  {\bf D} above) of using
  finestructure conastant data and our quantum correction to SU(5)
  model. So we must consider the agreement of the straight line story very good.

%30\%
  \section{Fine structure constants}
  \label{Finestructure}
We shall now review\cite{AppSU5} an article, in which I got the three fine
structure constants in the Standard Model written in terms of three
parameters, which all can be given a physical meaning and be calculated from
theory. This work is reminiscent of an SU(5) unified theory, but this SU(5)
symmetry is not a true symmetry of the theory of this work, which rather only
has gauge symmetry under the Standard Model Group $S(U(2)\times U(3))$ or even
more correct the third power of this group $(S(U(2)\times U(3))^3 =
S(U(2)\times U(3))\times S(U(2)\times U(3))\times S(U(2)\times U(3))$.
The $SU(5)$ like symmetry approximately for the fine structure constants only
comes in, because the action taken for the standard model group happens
in the first classical approximation to give the SU(5) relations between the
fine structure constants. Then the quantum correction is given as one of our
three
parameters due the  three seperate standard model groups it should be
multiplied by three compared to what it would be with only one standard
model group, but it can be calculated;
using the breaking to the diagonal subgroup of the three standard model
groups leads to  that the corrections for the inverse fine structure
constants for the diagonal subgroups becomes three times this first
calculated.
%43\%

The parameter $\mu_u$ giving
energy scale for the approximate SU(5) like synnetry is one of the energy
scales in the series of energy scales treated as the main point of the present
article, and can as such be determined from the other scales on our straight
line, and that is what we mean by, that this parameter is theoretically
calculable.

Let us, however, start by telling a tiny progress concerning the third of
our theoretically calculable parameters, namely the replacement for the
unified coupling of SU(5), for  which we have postulated the inverse
fine structure constant to be just three times the - on a lattice written
Standard Model group gauge theory - critical coupling for some expected
phase transition (or potentially instead a strong cross over).

\subsection{Critial coupling for Standard Model Group}

In our previous work\cite{AppSU5} we calculated from the experimentally
known fine structure constants (at say the Z-mass scale) and our form of
the expressions for the quantum corrections providing the deviations
from usual GUT-SU(5) fine structure constant relations the so to speak
experimental fine structure constants at the approximate unified scale:

Using the ratios of the deviations between the inverse finestrucuture
constants for the three standard model subgroups, U(1), SU(2), and SU(3)
predicted by our quantum correction model we found from the experimentally
known fine structure constants at say the $Z^0$-mass scale, the three
inverse running finestructure constants at the scale $\mu_u$ at which they
could meet in the sense of fitting our deviation ratios;
%From usual GUT-SU(5) fine structure constant relations the so to speak
%experimental fine structure constants at the approximate unified scale:
\begin{eqnarray}
  1/\alpha_3(\mu_u)|_{``data''}&=& 38.59\\
  1/\alpha_{1 \; SU(5)}(\mu_u)|_{``data''}&=& 41.43\\
  1/\alpha_2(\mu_u)|_{``data''}&=& 43.21
\end{eqnarray}

We suppose that the best way to extract from these due to the quantum
fluctuation effects from genuine SU(5) GUT deviating fine structure constants
at the approximate
unifying scale $\mu_u=5.12*10^{13}GeV$ an average to be compared with our
prediction of it being a critical value (seperating phases) just multiplied
by three (for the 3 one makes use of (\ref{diag})), is to use the with
dimension of the group
logarithmically weighted average.

%of the inverse of this average shall be just three times the
%critical (i.e. phase transition value) value for the inverted finestructure
%for the Standard Model, is to use the with dimension of the group
%weighted average.
The concept of seperate critical finestructure constant for the
three subgroups $U(1)$, SU(2) and SU(3) is not quite good, because one
shall really think of an a priori complicated phase diagram (see the figure
fromDon Bennetts thesis\cite{Donthesis}), in which
one for any combination of three fine structure constants have one phase
or the other realized. Indeed there are several different phases at least in
our mean field approximation used for getting an overview of the phases.
Among the different phase borders appearing according to this mean field
approximation we find in the thesis by Don Bennett\cite{Donthesis}, that
there is one border
separating one phase with confinement for all three simple subgroups
and one where they are approximately realized in an approximate perturbative
Yang Mills, although they or some of them might after all confine at lower
energy scales. This phase border corresponds to a transition at
a fixed value of  the simple subgroup dimension weighted
average of the logarithms of the three (inverse) finestructure constants.
Thus we believe that it is for this dimension of groups weighted average
that our - from old time suggested\cite{RDrel1,RDrel2, RDrel3, Don93, Don137}
- assumption that the ``fundamental''
fine structure constants should be critical can be applied.

%This latter concept of a critical value for a
%standard model coupling is not really such a good concept in as far as the
%Standard Model has three fine structure constants.
%However,we shall in a
%short moment estimating such a Standard model critical one by first looking up
%a critical coupling for an SU(5) group and then correcting it slightly
%to become
%one for the Standard Model group.
In fact we already said, that we shall find in Don Bennetts
thesis \cite{Donthesis}, that there is
a phase transition surface in the space of (inverse) finestrucutre constants,
which precisely lies along a surface, where the dimensionally weihgted
logarithms of the fine structure constant takes a specific value.

%Of course the ``correction'' cannot
%really be better than to allow the differences between the finestructure
%constants, but we anyway seek to do the best by using it for the dimension
%weighted average.

This dimension of group weighted average is for the experimental values
of the finestructure at the scale $\mu_u$, 
\begin{eqnarray}
  \hbox{Just dimension weighted: }&&\nonumber\\ 
  1/\alpha_{SMG}(\mu_u)|_{d-av.}&=& \frac{8}{12}*38.59
  + \frac{1}{12}41.43 + \frac{3}{12}*43.21\\
  &=& 25.73+ 3.45+ 10.80\\
  &=& 39.98;\\
  \hbox{Logarithmic averaging: }&&\nonumber\\
  log (1/\alpha_{SMG}(\mu_u))|_{l-d-av.}&=& \frac{8}{12}*log( 38.59)
  + \frac{1}{12}log(41.43) + \frac{3}{12}*log(43.21)\nonumber\\
  &=& 1.6013\\
  \hbox{giving }1/\alpha_{SMG}(\mu_u)|_{l-d-av.}&=& 39.93.
\end{eqnarray}
%50\%
Now we shall compare this $39.98$, which is kind of from data determined
unifed inverted finestructure constant with the critical coupling first for
$SU(5)$, which we got in the old work\cite{LR, LRN} to be
\begin{eqnarray}
  3/\alpha_{5\; crit}&=& 45.93\\
\end{eqnarray}
and below with our now believed to be better estimate of the SMG critical
coupling we shall get $38.2965 \pm 3$ (see equation (\ref{eq112})).

  \subsection{The Way from Don Bennetts Thesis}

  Actually using a crude mean field approximaton for estimating the phase
  diagrams for non-simple groups such as the Standard Model group Don Bennett
  in his thesis \cite{Donthesis} (see also \cite{Don93}) found the phase
  diagram for e.g. the standard model {\em group} as a function of what
  on the figure is denoted as the logarithms of the various simple groups,
  which together form the standard model group. Since the group volume
  is measured in the Cartan-Killing metric\cite{CK}, and the normalization
  of the
  latter is given by the fine structure constant for the simple group in
  question, you can in drawing phase diagrams use e.g. the logarithms of the
  group volumes to represent the fine structure constants. In the mean field
  approximation used in this thesis by Don Bennett it is so simple that the
  phase transition for a simple group occurs just, when the volume $Vol(G)$
  (normalized by the fine strucutre constant variable in terms of which we
  ask for a critical coupling) equals to $\sqrt{6\pi}$ rised to the power
  of the group dimension:
  \begin{eqnarray}
    Vol(G)_{crit}&=& \sqrt{6\pi}^{dim G}
  \end{eqnarray}

  For a cross product of just some simple groups such as
  $U(1)\times SU(2) \times SU(3)$ the phase diagram would
  just be a trivial cross product of the phase diagrams
  of the groups in the cross product with no interaction
  so to speak between the factors, because the volume
  of the cross product group is simply a product of that of the factors in
  the cross product.
  
  %The simple groups from which
  However, for e.g. the standard model group
  \begin{eqnarray}
    SMG = S(U(2)\times U(3)) &=&
    U(1)\times SU(2) \times SU(3)/
    \{(2\pi, -{\bf 1},\exp(i\frac{2\pi}{3}){\bf 1})^n |n \in {\bf Z}\}\nonumber
  \end{eqnarray}
  fluctuation in one simple group factor can influence the other
  factors and the phases of the system thus become
  %in a
  more complicated.
  %way.
  But we shall still assume, that it is group volume being on one or the
  other side of a certain value (not necessarily exactly
  $(6\pi)^{dim(G)/2}$)
  that matters for the phase. 
  %is composed would give a trivial in which there would be no
  %interactions between the various simpel groups if they were just
  %composed as a cross products as groups as they really are as Lie algebras.
  %But here is
  The complication/``interaction'' that one has divided out a subgroup
  $ \{(2\pi, -{\bf 1},\exp(i\frac{2\pi}{3}){\bf 1})^n |n \in {\bf Z}\} $ of
  the center of the cross product makes fluctuations in the fields
  for one simple group enhance the fluctuations of the by the division
  related simple group. We think of  that as 
  %one causes
  an 
  ``interaction'' between the simple subgroups. For this kind of combining
  the simple groups with their centers getting ``mixed up'' it is crucial
  that these simple groups have non-trivial centers, so let us remind the
  reader, that an $SU(N)$ group has a center consisting of the matrices
  \begin{eqnarray}
    Center(SU(N)) &=& \{ \exp(\frac{i k 2\pi}{N})*{\bf 1}_{N\times N}|
    k\in {\bf Z }\}.
  \end{eqnarray}
  %57\%
  One can thus corresponding to each $SU(N)$ form one with its center
  divided out
  \begin{eqnarray}
    SU(N)_{with \; center\; divided \; out}&=& SU(N)/
    \{ \exp(\frac{i k 2\pi}{N})*{\bf 1}_{N\times N}|k\in {\bf }\}
    \end{eqnarray}
  To divide out the center of course diminishes the volume by a factor
  $N$, and thus according to the rule (in mean field approximation)
  that the critical coupling correspond to the volume being
  $\sqrt{6\pi}^{dim G}$ one would have to adjust the fine structure
  constant to again bring the volume with the new critical
  coupling to be again $\sqrt{6\pi}^{dim G}$. So
  \begin{eqnarray}
    \frac{1}{\alpha_{N\;  crit}}|_{with \; center\; divided \; out}&=&
    N^{\frac{1}{dim( SU(N))/2}}\frac{1}{\alpha_{N \,crit}}.
  \end{eqnarray}

\includegraphics[scale=0.7]{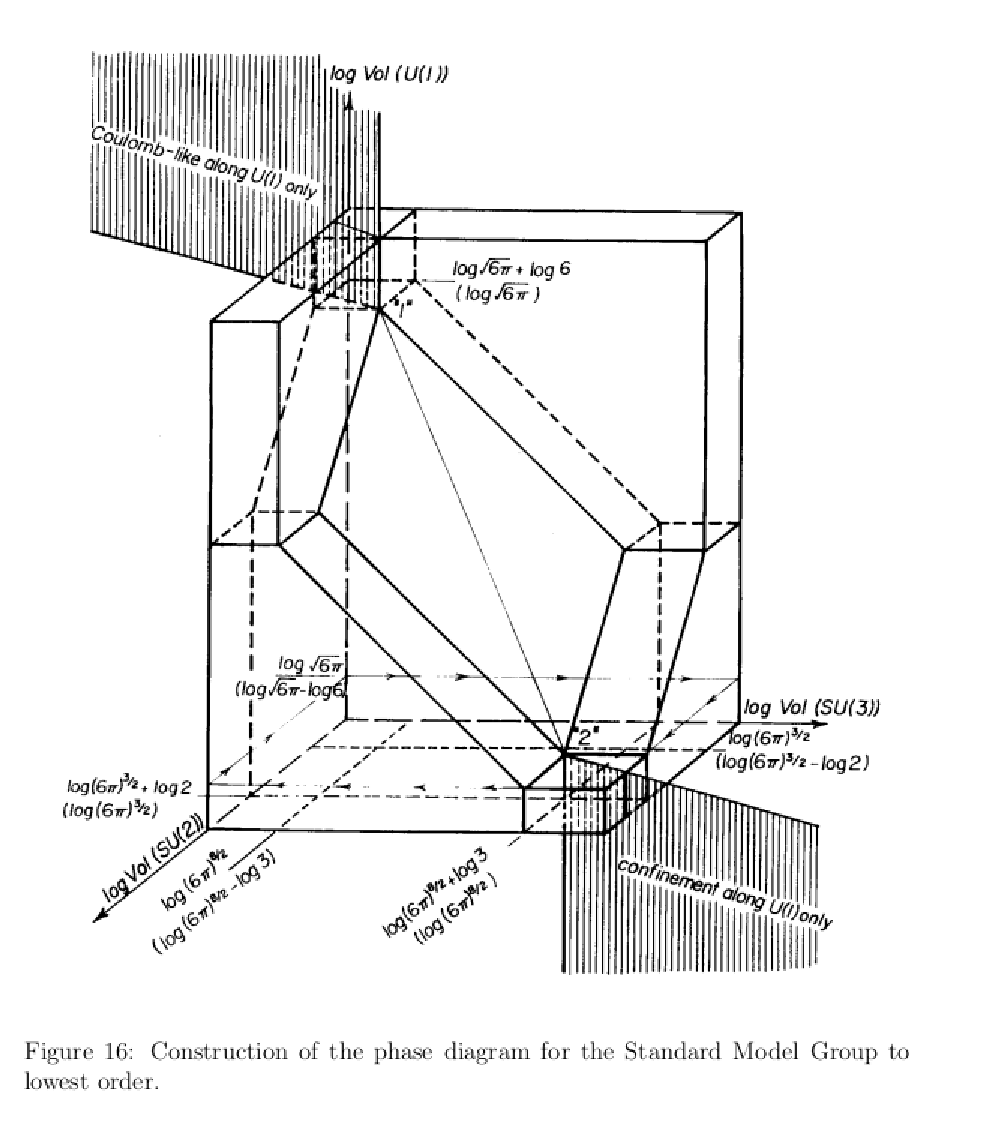}

\subsection{Critical Fine structure constant for the Standard Model Group}

The mean field approximation used by Don Bennett suggests, that for a group
that is a simple Cartesian product of some groups, as e.g. the Standard Model
Group without the division out of the part of the center otherwise
being in the definition of this group, the phase transition border
or just transition point determined for the product
\begin{eqnarray}
  1/\alpha_{1\; SU(5)}*(1/\alpha_2)^3*(1/\alpha_3)^8&having& ``special \; value''
  \label{pr}\\
&  \approx& ``related \; to'' (6\pi)^{12}.
\end{eqnarray}
When we in the same approximation want the critical finestructure constants
at the phase transition between the confinement phase with total
``confinement'' for the full group with some subset of the center divided out,
as e.g. the true Standard model group $G_{SMG}=S(U(2)\times U(3))$, we shall
just calculate the factor $Cd(Z_{do})$ = ``the cardinal number
of the subsgroup of the center being divided out'', by which the volume
of the group is being reduced by this division out, and then the
dimension weighted product of the inverse fine structure constants
\ref{pr} shall get its for criticallity required value increased by this
factor. So e.g. this product is for the Standard Model Group
$G_{SMG}=(U(1)\times SU(2)\times SU(3))/{\bf Z}_6$ a factor 6 larger than for
the simple Cartesian product
of the three groups from which the Standard Model one is composed. I.e.
\begin{eqnarray}
  [1/\alpha_{1\; SU(5)}*(1/\alpha_2)^3*(1/\alpha_3)^8]_{crit\; for \; G_{smg}}&=&
 6* [1/\alpha_{1\; SU(5)}*(1/\alpha_2)^3*(1/\alpha_3)^8]_{crit \; 
   cross\, product}\nonumber\\
 &=&  6* [1/\alpha_{1\; SU(5)}*(1/\alpha_2)^3*(1/\alpha_3)^8]_{crit \; 
   U(1)\times SU(2)\times SU(3)}.\nonumber
  \end{eqnarray}
%54\%
This form is the reason, why the combination of the three
fine structure constants, which can be postulated to have a critical
vaue - or as we shall assume in our model just 3 times the critiacle
value - is the logarithmically dimension weighted average of the inverse
fine structure
constants, as explained more in subsection \ref{av}.

\subsection{Calculation of the Critial (Inverse) Finestructure constant
  average}

For estimating the critical coupling for the Standard Model {\bf group}
SMG we shall make use of our formula by Laperashvili et al. developped
by use of renorm group and monopoles being assumed and used for the
critical inverse coupling squares ($\sim$ inverse fine structure
constants)
\begin{eqnarray}
  \alpha^{-1}_{N \, crit}&=& \frac{N}{2}\sqrt{\frac{N+1}{N-1}}
  \alpha^{-1}_{u(1)\; crit}\\
  \hbox{where }\alpha_{U(1) \; crit}^{lat}&\approx& 0.20\pm 0.015\\
  \hbox{or } \alpha_{U(1) \; crit}^{-1} &\approx & 5 \pm 7.5 \% =5 \pm 0.4 
\end{eqnarray}
%63\%

\subsection{The Average, that can be Calculated as Critical}
\label{av}

In \cite{AppSU5} we find, that insisting on the ratios of the differences
of the inverse finestructure constants at the ``approximate unification ''
should be as required from our quantum correction model - i.e. the
$1/\alpha_{1\; SU(5)}$ shall divide the interval from $1/\alpha_3$ to
$1/\alpha_2$ into pieces in the ratio 3:2 - the `` unification scale
inverse finestructure constant'' would have to have the inverse
fine structure constants
\begin{eqnarray}
  1/\alpha_{1\; SU(5)}(\mu_u) &=& 41.355\pm 0.017\\
  1/\alpha_2(\mu_u) &=& 43.203\pm 0.02\\
  1/\alpha_3(\mu_u) &=& 38.585\pm 0.05.
\end{eqnarray}
when using the experimental fine structure constants
(we have reproduced these ``experimental'' fine sturcture
constants below in (\ref{exp1}, \ref{exp2}, \ref{exp3}) in this article)
at say the $M_Z$ mass.
%59\%

Because it is supposed to be the total group volume that matters it is
expected, that it is the logarithmic average of these quantities weighted
by the dimensions of the simple groups that we shall expect to be
obtainable as critical value:

The basis 10 logarithms for these inverse fine structure constants
at our replacement for unification scale $\mu_u$ are
\begin{eqnarray}
  log(1/\alpha_{1\; SU(5)}(\mu_u)) &=&  1.6165\\
  log(1/\alpha_2(\mu_u))&=&1.6355\\
  log(1/\alpha_3(\mu_u))&=& 1.5864\\
  \hbox{Average } log(1/\alpha_{av}(\mu_u)) &=& \frac{1*1.6165 + 3*1.6355+
    8*1.5864}{12}\\
  &=& \frac{19.2143}{12}\\
  &=& 1.6012\\
  \hbox{giving } 1/\alpha_{av}(\mu_u)&=& 39.920.\label{expav} 
\end{eqnarray}

\subsection{Calculation of Critical Coupling}
% Here some trouble because some capital letters do not work
% on my keyboard.
Using these formulas we get
\begin{eqnarray}
  &\hbox{SU(2)}(/Z_2)&\\
  1/\alpha_{crit\, SU(2)}&=& 1/\alpha_{1\, crit}* \frac{2}{2}*
  \sqrt{\frac{2+1}{2-1}}\\
  &=&
  1/\alpha_{U(1)\, crit} * \sqrt{3}= 8.660\\
  \hbox{while }1/\alpha_{SU(2)/z_{2\, crit}}=1/\alpha_{SO(3)\, crit}&=& 1/\alpha_{U(1)\, crit}\sqrt{3}
  * \sqrt[3]{4}=13.747\nonumber
  \\
  \hbox{to compare with } \hbox{``old'' }  1/\alpha_{2\, crit}&=&15.7\pm 1
  \hbox{(unexponentiated)}\nonumber \\
  &=& 16.5\pm 1 \hbox{(exponentiated)}\\
  &\hbox{SU(3)}(/Z_3)&\nonumber\\
  \hbox{``old'' } 1/\alpha_{SU(3)\; crit}&=& 1/\alpha_{U(1)\; crit}*
  \frac{3}{2}\sqrt{\frac{3+1}{3-1}}\\
  &=& 5*\frac{3}{\sqrt{2}}=5*2.1213\\
  &=& 10.6066\\
  \hbox{while } 1/\alpha_{SU(3)/Z_3 \; crit}&=&1/\alpha_{U(1)\; crit}*
  \frac{3}{\sqrt{2}}*3^{2/8}\\
  &=& 13.9591\\
  \hbox{to compare with } \hbox{``old''} 1/\alpha_{3\; crit}&=&17.7\pm 1
  \hbox{(unexponentiated)}\nonumber\\
  &=& 18.9 \pm 1 \hbox{(exponentiated)}
\end{eqnarray}
We here compared with our old estimates, see e.g. Don Bennett's
thesis\cite{Donthesis},
%60\%
from which we have included as figures the most cricial formulas, and find
that our expressions using the $1/\alpha_{SU(N)\; crit} = \frac{N}{2}*
\sqrt{\frac{N+1}{N-1}}$ \cite{Ryzhikh} tend to give a bit lower inverse
critical fine structure constants than the ``old'' estimates. These ``old''
critical inverse fine structure constant estimates were truly lattice artifact
calculations made for a meeting of three phases - a confining one, an
approximate one with $SU(N)/Z_N$ surviving as approximately pertubative
Yang Mills, and finally one where it is SU(N), that suvives -, and so the
monopoles in these ``old'' estimates were lattice-artifact monopoles. Contrary
our presently used calculation\cite{Ryzhikh, LR,LRN} rather has a monopole
%61\%
in the continuum -which is
better in agreement with the very speculative story in the present
work, that a resonance decaying into a pair of muons of mass 27 GeV should
be related to monopoles -. The reason we suggest, that it is the critical
couplings for the SU(2) and SU(3) with their center divided out, so
really the critical couplings for $SU(2)/Z_2$ and $SU(3)/Z_3$, that should
be compared to the ``old'' numbers, is that the dominant term in the
lattice theory for these critical couplings are the ones corresponding
to these groups with the center divided out.

The distingtion ``unexponentiated'' versus ``exponentiated'' is just
a tiny variation in the corrections in the ``old'' calculation, see 
Don Bennetts thesis\cite{Donthesis}.
%61\%

\begin{adjustbox}{width = \textwidth}
\includegraphics[scale=1.0]{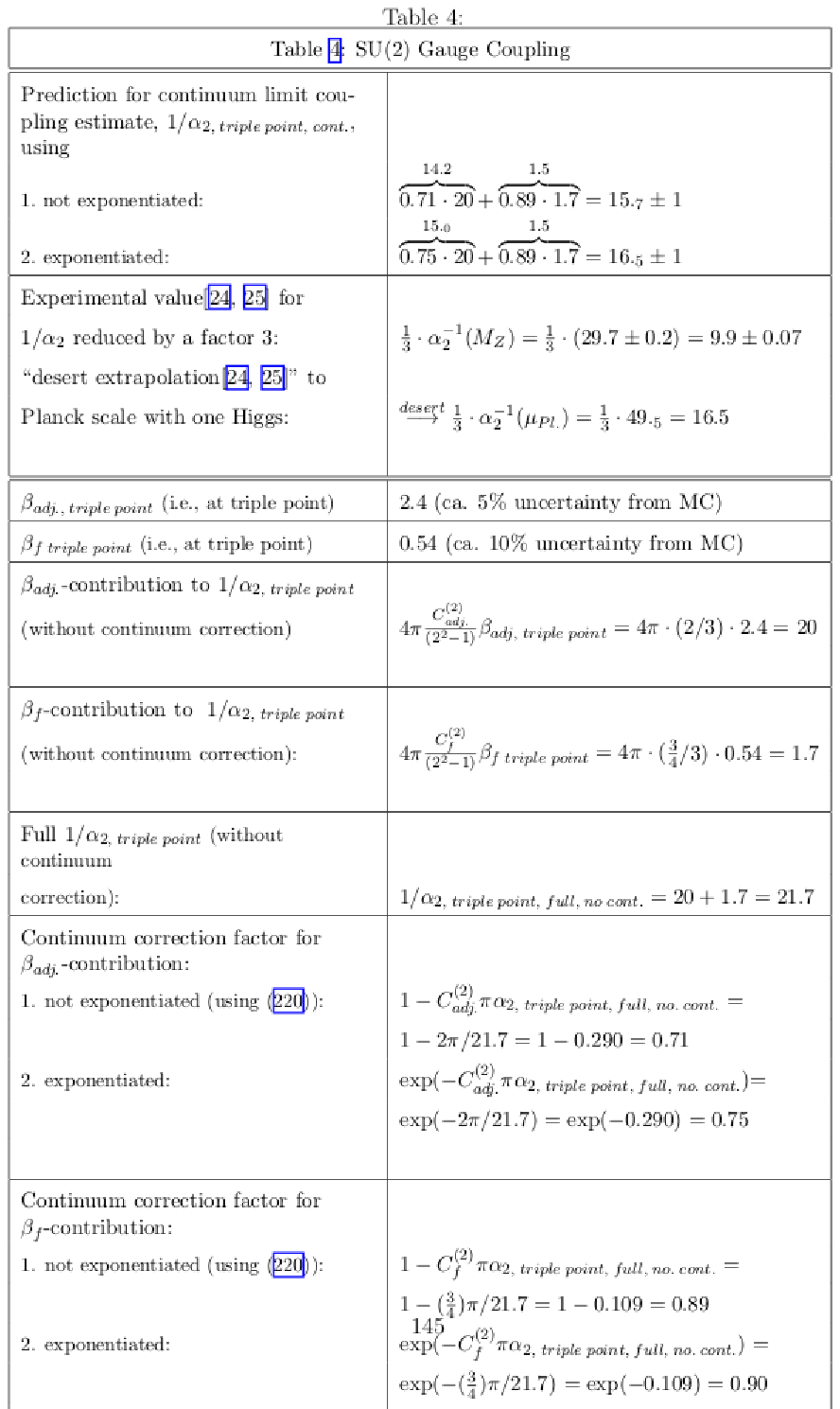}
\end{adjustbox}

\begin{adjustbox}{width=\textwidth}
\includegraphics[scale=0.9]{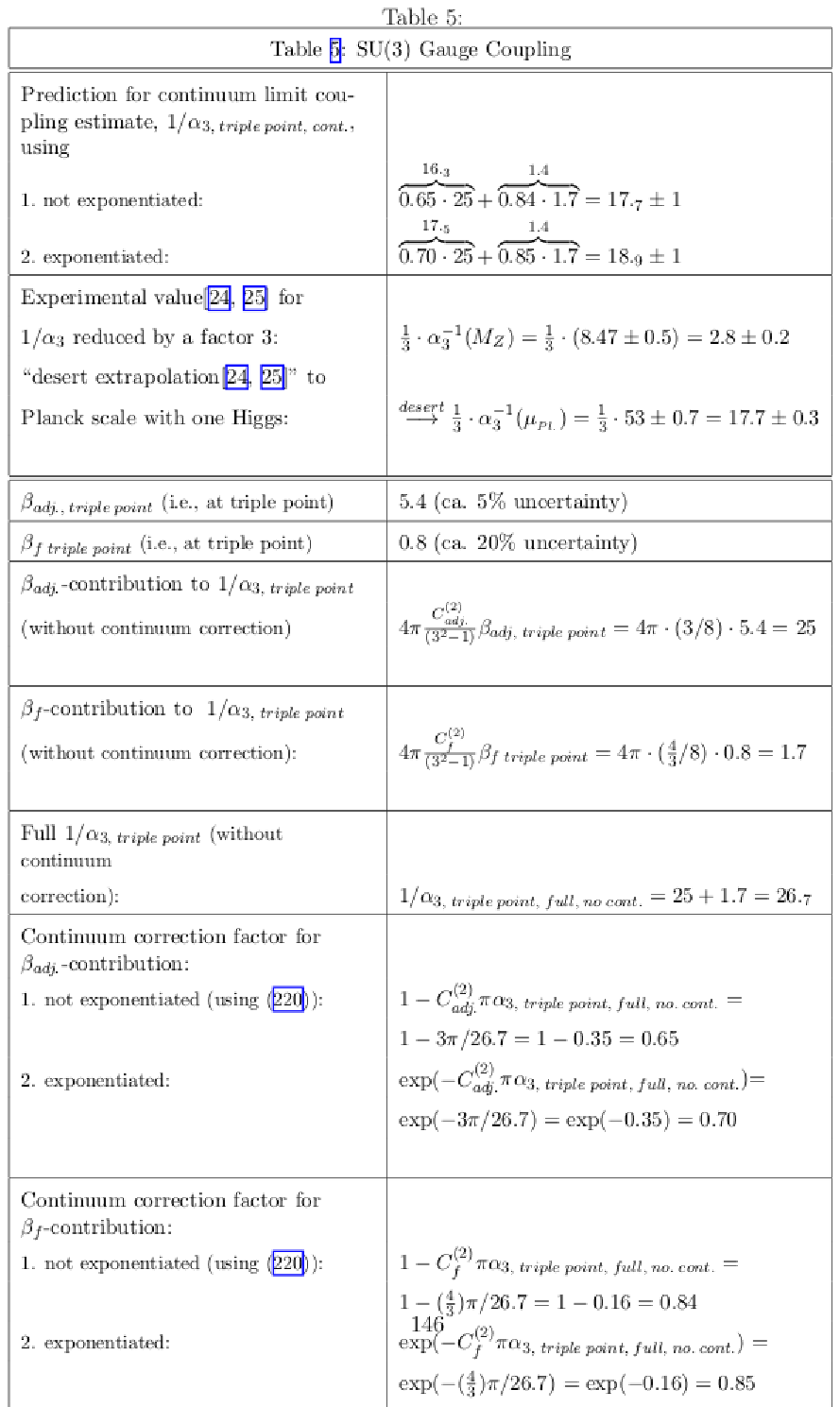}
\end{adjustbox}
% test: 
%51\%

\subsubsection{Our Expression for the SMG critical Inverse Fine structure
  constant,
  %reachable averrage
}

Let us now use our formulas to get first a critical couplig relation
for the simple cross product group $U(1)\times SU(2)\times SU(3)$ and 
then for the Standard odel {\bf group} $SMG= S(U(2)\times U(3))$ for the
logarithmically dimension weighted (inverse) fine structure constant:

\begin{eqnarray}
  &&U(1)\times SU(2) \times SU(3)\\
  1/\alpha_{av \; U(1)\times SU(2)\times SU(3)\; crit}&=&\sqrt[12]
  {1/\alpha_{U(1)\; crit})*
    (1/\alpha_{SU(2)\; crit})^3*(1/\alpha_{SU(3)\; crit})^8}\nonumber\\
  &=& \sqrt[12]{(1/\alpha_{U(1)\; crit})^{12}*(\sqrt{3})^3*(\frac{3}{2}*\sqrt{2})^8}
  \\
  &=& 1/\alpha_{U(1)\; crit}*\sqrt[12]{3^{3/2}*3^8*2^{-4}}\\
  &=& 1/\alpha_{U(1)\; crit}*3^{19/24}*2^{-1/3}\\
  &=& 5*2.3863/1.2599\\
  &=& 9.4699\\
  &&SMG =S(U(2)\times U(3))\nonumber\\
  1/\alpha_{av \; SMG\; crit}&=& \sqrt[12]{6*1/\alpha_{U(1)\; crit}*
    (1/\alpha_{SU(2)\; crit})^3*(1/\alpha_{SU(3)\; crit})^8}\nonumber\\
  &=& \sqrt[12]{6^2
    *1/(\alpha_{U(1)\; crit})^{12}*(\sqrt{3})^3*(\frac{3}{2}*
    \sqrt{2})^8}\\
  &=& 1/\alpha_{U(1)\; crit}6^{2/12}*3^{19/24}*2^{-1/3}\\
  &=& 1/\alpha_{U(1)\; crit}*3^{23/24}*2^{-1/6}\\
  &=&5*2.86577 *0.890899\\
  &=& 5*2.55311\\
  &=& 12.76550\\
  3/\alpha_{av \; SMG\; crit}&=&38.2968\label{critt3}
\end{eqnarray}

This last number (\ref{critt3}) and  the experimental
averaged number $\frac{3}{\alpha_{av}(\mu_u)}$, (\ref{expav}), agrees with a
deviation
of only 1.6 (meaning only 4\% ) which is even agreement with minimal
estimate of the
uncertainty of the theoretical number of 7.5 \%, which would mean one standard
deviation being 2.9.

\subsection{Resume of the Fine strucutre calculation}

We can resume our model for the fine structure constants by writting
the values of the inverse fine structure constants for the three subgroups
U(1), Su(2), amd SU(3) in SU(5)-adjusted notation (it just means we use
$1/\alpha_{1 \; SU(5)}=\frac{3}{5}*1/\alpha_1$ instead of the $1/\alpha_1$ itself
for the Standard Model U(1)) in terms of the three parameters
$\ln(\frac{\mu_u}{M_Z})$, $q$, and $3/\alpha_{av \; SMG \; crit}$, which
we claim, we have calculated, under our assumptions of course:
\begin{eqnarray}
  1/\alpha_{1 \; SU(5)}(M_Z) &=&\frac{\frac{41}{10}}{2\pi}*\ln(\frac{\mu_u}{M_Z})
  +\frac{3}{10}*q +\frac{3}{\alpha_{av \; SMG \; crit}}\\
  1/\alpha_{2}(M_Z)&=& \frac{-\frac{19}{6}}{2\pi}*\ln(\frac{\mu_u}{M_Z})+
  \frac{7}{10}*q+ +\frac{3}{\alpha_{av \; SMG \; crit}}\\
  1/\alpha_{3}(M_Z)&=&\frac{-7}{2\pi}*\ln(\frac{\mu_u}{M_Z})-\frac{3}{10}*q+
   \frac{3}{\alpha_{av \; SMG \; crit}}\\
  \end{eqnarray}

The terms with $q$ give the deviation from genuine SU(5) GUT and represent
the quantum correction in the  lattice theory to an action, which in the
classical approximation
happens to give the  SU(5) invariance relation betwen the
three standard model fine strucutre constants. The action we
imagine in our model is namely expressed in terms of a $5\times 5$
marix representation of the only true gauge theory in our model - that of the
Standard Model group -, and the simplest trace happens to be identical to
an SU(5) action. The coefficients to $q$, i.e.  3/10, 7/10, and -3/10, have
been arranged, so that the correction to the difference
$1/\alpha_2 -1/\alpha_{1 \; SU(5)}$ becomes 2/3 of that of the difference
$1/\alpha_{1\; SU(5)}-1/\alpha_3$
  as estimated in our article \cite{AppSU5}.
  Further they are arranged to make contribution to the dimension weighted
  average zero, i.e.
  \begin{eqnarray}
    1*\frac{3}{10} + 3*\frac{7}{10}+ 8*\frac{-3}{10} &=& 0.
    \end{eqnarray}
  The parameter $q$ would, if there was only a simple Wilson lattice
  (in one layer) be $q=\pi/2$, but it is an important physical assumption
  in our model, that we do not truly have only the Standard Model
  gauge group, but rather a {\bf cross product of three} Standard Model groups
  with each other together. This makes the quantum correction three times
  as big (see (\ref{diag})), and we thus have
  \begin{eqnarray}
  q&=& 3*\pi/2 =  4.7124. 
  \end{eqnarray}

  The same factor 3 signaling, that the genuine gauge grouop should be
  $SMG\times SMG \times SMG$ rather than just $SMG$, comes in and makes
  averaged inverse finestructure constant at the essential unification scale
  $\mu_u$ become $3/\alpha_{av \; SMG \; crit}$ rather than just the critical
  inverse finestructure constant itself. It is supposed that this factor
  3 is the number of families. So to speak: Earch family has its own SMG gauge
  group.

  Let us collect the {\bf Theoretical Parameter Values}:

  Let us first assign uncertainties which are minimal needed, i.e. it would be 
  hard to avoid these uncertainties, but there might be further ones
  (e.g. our formula for critical coupling\cite{LRN, LR} could have similar
  or further
  uncertainties):
  \begin{eqnarray}
    \ln{\mu_u}{M_Z}=\ln(\frac{4.9534*10^{13}GeV}
       {91.1876 GeV})&=&27.0204\pm 0.01
       \hbox{(see (\ref{lnmuuomz})) }\\
       q=3*\pi/2 &=& 4.7124 \pm 0.05\\
       \frac{3}{\alpha_{av\; SMG \; crit}} =3*12.76550&=&38.2965\pm 3 \label{eq112}
  \end{eqnarray}

  (The error on q we propose to take as a higher order correction
  and thus percentwise of the order of $\alpha$.)

  We can now simply calculate the predictions for the inverse fine structure
  constants at the $M_Z$ scale and compare with the experimentally determined
  values:
  \begin{eqnarray}
    1/\alpha_{1\; SU(5)}(M_Z)|_{predicted} &=&
    \frac{\frac{41}{10}}{2\pi}*27.0204
    +\frac{3}{10}*4.7124 +38.2965\nonumber\\
    &=& 17.6318 +1.4137+38.2965\\
    &=& 57.3420\pm 3\\
    \hbox{to compare with }1/\alpha_{1\; SU(U)}(M_Z)|_{exp}&=& 59.008\pm 0.013.\\
    1/\alpha_{2}(M_Z)&=& \frac{-\frac{19}{6}}{2\pi}*27.0204+
    \frac{7}{10}*4.7124 +38.2965\nonumber\\
    &=& -13.6180+3.2987+38.2965\\
    &=& 27.9772\pm 3\\
    \hbox{to  compare with } 1/\alpha_2(M_Z)|_{exp}&=& 29.569\pm 0.017\\
  1/\alpha_{3}(M_Z)&=&\frac{-7}{2\pi}*27.0204-\frac{3}{10}*4.7124+
  38.2965\nonumber\\
  &=& -30.1030-1.4137+38.2965\\
  &=& 6.7798\pm 3\\
  \hbox{to compare with } 1/\alpha_3(M_Z)|_{exp}&=& 8.446\pm 0.05
  \end{eqnarray}

  Our predictions agree wonderfully within the by the critical coupling
  parameter $\frac{3}{\alpha_{av\; SMG\; crit}}$ dominated uncertainty, which
  we had put to $\pm 3$. However the deviation is very systematic, all the
  three inverse fine structure constants being just $1.6414$ bigger
  experimentally than our prediction. So if we gave up trusting accurately
  our critical coupling parameter  $\frac{3}{\alpha_{av\; SMG\; crit}}$,
  and instead just fitted it to the experimental data, while still keeping
  our two other theoretically predited parameters, $q$ and
  $\ln(\frac{\mu_u}{M_Z})$,
  %as our theory predict,
  we could hope for a higher
  accuracy predition. In fact let us change our critical coupling
  parameter to a to data fitted value by replacing it like:
  \begin{eqnarray}
    \frac{3}{\alpha_{av\; SMG\; crit}}=38.2965 &\rightarrow& 38.2965+1.6414=
    39.9379
  \end{eqnarray}

  Then we would get rather:
 \begin{eqnarray}
    1/\alpha_{1\; SU(5)}(M_Z)|_{predicted} &=&
    \frac{\frac{41}{10}}{2\pi}*27.0204
    +\frac{3}{10}*4.7124 +39.9379\nonumber\\
    &=& 17.6318 +1.4137+39.9379\\
    &=& 58.9834\pm 0.02\\
    \hbox{to compare with }1/\alpha_{1\; SU(U)}(M_Z)|_{exp}&=& 59.008\pm 0.013.
    \label{exp1}\\
    1/\alpha_{2}(M_Z)&=& \frac{-\frac{19}{6}}{2\pi}*27.0204+
    \frac{7}{10}*4.7124 +39.9379\nonumber\\
    &=& -13.6180+3.2987+39.9379\\
    &=& 29.6186\pm 0.04\\
    \hbox{to  compare with } 1/\alpha_2(M_Z)|_{exp}&=& 29.569\pm 0.017
    \label{exp2}\\
  1/\alpha_{3}(M_Z)&=&\frac{-7}{2\pi}*27.0204-\frac{3}{10}*4.7124+
  39.9379\nonumber\\
  &=& -30.1030-1.4137+39.9379\\
  &=& 8.4212\pm 0.02\\
  \hbox{to compare with } 1/\alpha_3(M_Z)|_{exp}&=& 8.446\pm 0.05\label{exp3}
  \end{eqnarray}

 The agreement is still within our a bit arbitrarily estimated uncertainties
 in spite of, that we now have the uncertainty on the second decimal in the
 inverse fine structure constants, and we only fitted {\bf one} of our
 {\bf three} theoretically predicted parameters! The $q$ being the quantum
 correction breaking SU(5), and the replacement for unification scale
 $\mu_u$ used in our parameter $\ln(\frac{\mu_u}{M_Z})$ seemingly are so
 accurate as to admit for only deviations on the second decimal after
 the ``.''! The difference $1/\alpha_{1\: SU(5)}(M_Z) - 1/\alpha_{2}(M_Z)$
 predicted to $29.365$ with one promille accuracy was experimentally
 $29.439$ deviating by only $0.074$ meaning 2.5 \permil.

 \section{Conclusion}
 \label{Conclusion}
 We have found a phenomenologically surprisingly well agreeing
 relation involving about 9 different energy scales in physics and the
 supposed power $n$ of the link variable $a$ which should be relvant for these
 different scales $a^n$. In fact the logarithms of the energies of the different
 scales versus the power $n$  of the link variable associated turned out
 to be a
 linear relation described by a straight line. It must be admitted that
 most of these energy scales are only meaningfull order of magnitudewise.
 But at the end we considered four of the scales, which at least had the
 chanse of giving more than only order of magnitude numbers for the energies.
 It turned out, that taking it, that we should use the energy occuring  in the
 Lagrangian (as a coefficient being this energy to some  power) these four
 points fall very well on the straight line with
 the accuracy achievable with them. In fact we used three of the points
 among these four to make a prediction for the one being the unification
 scale for an approximate SU(5) unification of the fine structure constants.
 Using this unification scale from the straight line $\mu_u$ together
 with our earlier model for a quantum effect breaking the SU(5), which was
 in our picture only an accidental classical approimation being SU(5)
 invariant, we obtained values for the differences between the
 three inverse fine strucutre constants in the Standard Model deviating
 only from the experimentally determined numbers corresponding to an error
 $\pm 0.05$ for the these differences, which are e.g. 21.20. So our
 model agrees for these differences to {\bf a quarter of a percent!} 

 We also did obtain predictions for the fine structure constants proper, but
 because of the dependence on the at laest so far not so well calculated
 critical fine structure constant for the Standard Model group (even the concept
 might be not well enough defined) we only get the inverse fine structure
 constants with error estimates of the order $\pm 3$, but even that is very
 good!

 Our story of the straight line for the logarithms of the energy scales
 versus the power to which the link size $a$ should be raised  to relate to
 the scale in question, is indeed very intriguing. For some of the energy
 scales we can easily imagine, that under the assumption of, the parameters
 of the lattice theory being of order unity compared to the local lattice links,
 we get the energy scale given by some power of the link, averaged of
 course appropriately. This is for instance the case for the scales as
 the approximate unification scale $\mu_u$ and the Planck scale, and scales
 for masses of a lot of bosons, or of a lot of fermions (taken to be the
 see-saw scale). But when we come to the hadron-string scale, then
 one would say, that properties of hadron physics should be given by
 QCD alone and not depend much on the lattice, since the lattice should
 function
 only as a cut off and when expressed in terms of the renormalized couplings,
 the cut off should drop out. Nevertheless it was most accurately the
 hadron string scale, we used in our so successful prediction of the
 inverse fine structure constant differences. So one would say, that it is
 physically absurd, that this hadron string scale should fall on the same line
 as the scales, which are  easy to conceive of as lattice link size connected.
 Either one
 would need to say, that something more than just QCD is involved in making
 confinement and some effective hadronic strings, or one would have to take
 it that some mysterious - yet to be understood - principle in the lattice
 physics just arrange this lattice to produces its string action to just
 agree with that of 
 the effective hadron strings. Both ways to explain the strange
 coincidence sounds physically very strange.

 The domain wall scale, which we expect would be due to some phase border
 in a mainly QCD and quark physics grounded physics, would be in the same
 way intriguing and mysterious. It would again require either something else
 than just QCD-physics, or some mysterious adjustment of the
 lattice knowledgeable about QCD, or may be opposite some QCD parameter
 should be adjusted to make our straight line coincidence work.

 The scale of monopole or related particle mass, could better without
 too much mystery be an effect of an ontological lattice.

 \subsection{Outlook}

 As an optimistic outlook we imagine the possibility, that our high
 accuracy success with the inverse fine structure constant differences
 could open the way for a numerological study of higher energies
 than what is directly reachable by the accellerators of the day.

 A point, that came out of the present works, is  that the cut off scale in
 our model is much closer, i.e. much
 lower in enrgy, than for instance, what many of us would have believed before,
 the Planck scale. Thus cut off effects are potentially to be seen in very
 accurately measured quantities, such as the anomalous magnetic moments, or
 the finestrucure constants themselves. In our Corfu 2024 proceedings\cite{Ko24}
 we gave an example of how to estimate such cut off effects in our model.
 The cut off effects seem to be on the borderline of being observable with the
 present experimental accuracy.

%\end{document}

\section*{Acknowledgements}
The author thanks the Niels Bohr Institute for status as emeritus.
This work was discussed in
both Bled Workshop 2024 and 2025 and the Corfu Insitute last year 2024.

%\reftitle{References}

\end{document}